 \documentclass[preprint2]{aastex}

\def\gax    {${_>\atop^{\sim}}$}
\def\aox    {$\alpha_{ox}$}
\def\etal   {{\it et al.}~}

\def\kms    {km s$^{-1}$}

\shorttitle{Mathur et al.}
\shortauthors{Thomson Thick Absorption in a BALQSO}

\begin{document}


\title{Thomson Thick X-ray Absorption in a Broad Absorption Line Quasar
PG0946+301 }


\author{S. Mathur\altaffilmark{1,2}, P. J. Green\altaffilmark{2},
N. Arav\altaffilmark{3},
M. Brotherton\altaffilmark{4},
M. Crenshaw\altaffilmark{5},
 M. deKool\altaffilmark{6},
M. Elvis\altaffilmark{2},
R. W. Goodrich\altaffilmark{7},
F. Hamann\altaffilmark{8},
 D. C. Hines\altaffilmark{9},
V. Kashyap\altaffilmark{2},
K. Korista\altaffilmark{10},
 B. M. Peterson\altaffilmark{1},
J. Shields\altaffilmark{11},
 I. Shlosman\altaffilmark{12},
 W. van Breugel\altaffilmark{13}
M. Voit\altaffilmark{14}
}





\altaffiltext{1}{The Ohio State Univ., smita@astronomy.ohio-state.edu}
\altaffiltext{2}{Harvard Smithsonian Center for Astrophysics}
\altaffiltext{3}{University of California, Berkeley}
\altaffiltext{4}{NOAO, Tucson, Arizona}
\altaffiltext{5}{Catholic
University of America and NASA's Goddard Space Flight Center}
\altaffiltext{6}{RSAA, ANU, Australia}
\altaffiltext{7}{The W. M. Keck Observatory, Hawaii}
\altaffiltext{8}{University of Florida}
\altaffiltext{9}{Steward Observatory, The University of Arizona}
\altaffiltext{10}{Western
Michigan University}
\altaffiltext{11}{Ohio University}
\altaffiltext{12}{University of Kentucky}
\altaffiltext{13}{LLNL}
\altaffiltext{14}{Space Telescope Science Institute}


\begin{abstract}

 We present a deep ASCA observation of a Broad
 Absorption Line Quasar (BALQSO) PG0946+301. The source was clearly detected
 in one of the gas imaging spectrometers, but not in any other
 detector. If BALQSOs have intrinsic X-ray spectra similar to normal
 radio-quiet quasars, our observations imply that there is Thomson
 thick X-ray absorption (N$_H$\gax$10^{24}$ cm$^{-2}$) toward
 PG0946+301. This is the largest column density estimated so far
 toward a BALQSO. The absorber must be at least  partially ionized and
 may be responsible for attenuation in the optical and UV. If the
 Thomson optical depth toward BALQSOs is close to one, as inferred here,
 then spectroscopy in hard X-rays with large telescopes like XMM would
 be feasible.

\end{abstract}




\section{Introduction}

 About 10 - 15\% of optically selected QSOs have optical/UV spectra
showing deep absorption troughs displaced blueward from the
corresponding emission lines.
These broad absorption lines (BALs)
are commonly attributed to material flowing toward the observer with
velocities of up to $\sim 50,000~$ \kms. BALQSOs are probably normal
QSOs viewed at a fortuitous orientation passing through a BAL outflow,
thus implying a BAL ``covering factor'' at least 10 - 15\% in all
QSOs.  BALQSOs thus provide a unique probe of conditions near the
nucleus of most QSOs. The absorbing columns typically inferred from
the UV spectra for the BAL clouds themselves are $N_{\rm H}\sim 10^{20-21}$
cm$^{-2}$
(Korista \etal 1993). It has been noted, however, that UV studies
underestimate the BAL column densities because of saturation (Korista
\etal 1993, Arav 1997, Hamann 1998). BALQSOs, as a class, show higher
optical/UV polarization than other radio-quiet QSOs (Schmidt \&
Hines 1999, Ogle \etal 1999). Polarization studies reveal multiple
lines of sight through high column density gas (Goodrich \& Miller
1995, Cohen \etal 1995).

 With the absorbing column densities as estimated from the earlier UV studies,
we
would have expected very little soft X-ray absorption in the BALQSOs.
However, BALQSOs are found to be markedly underluminous in X-rays
compared to their non-BALQSO counterparts (Bregman 1984, Singh \etal
1987, Green et al. 1995). Green \& Mathur (1996, here after GM96)
studied 11 BALQSOs observed with ROSAT and found that just one was
detected with \aox \footnote{The slope of a hypothetical power law
connecting 2500\,\AA~ and 2~keV is defined as \aox\, = $0.384~{\rm
log} L_{opt}/L_x $, so that \aox\, is larger for objects with weaker
X-ray emission relative to optical.} about 2.  BALQSOs thus have unusually weak
soft X-ray emission, as evidenced by large \aox (\gax
1.9. c.f. \aox=1.51$\pm0.01$, from Laor \etal 1997, for radio-quiet
quasars). If BALQSOs are indeed normal radio-quiet QSOs, then their
weak X-ray flux is most likely due to strong
absorption. Unfortunately, due to the low observed flux, there are no
observed X-ray spectra of BALQSOs to confirm the absorption scenario,
with one exception, the archetype BALQSO PHL5200 (Mathur, Elvis \&
Singh 1995, here after MES95). The ASCA spectrum of PHL5200 is best
fit by a power-law typical for non-BALQSOs in the 2--10 keV range,
with intrinsic absorption 2 to 3 orders of magnitude higher than
inferred from UV spectra alone (MES95). However, the PHL5200 spectrum
suffers from a low signal to noise ratio, and while the above was a
preferred fit, a model with no intrinsic absorption also fits the
data. Recently Gallagher \etal (1999, hereafter G99) studied a sample
of six new BALQSOs with ASCA, of which two were detected. G99 derived
column densities of \gax 5$\times 10^{23}$ cm$^{-2}$ to explain the
non-detections, even higher than the ROSAT estimates (assuming a neutral
absorber with solar abundances unless stated otherwise).

How are the X-ray and UV absorbers related to each other? Are they
both part of the same outflow?  If so, then the kinetic energy carried
out is a significant fraction of bolometric luminosity of the quasar
(see Mathur, Elvis \& Wilkes 1995 for a discussion). With all QSOs
likely to contain a BAL outflow, it becomes very important to measure the
absorbing column density accurately to understand the energetics and
dynamics of quasars. We attempt this  with a deep ASCA observation
of a typical BALQSO, PG0946+301.

\section{Observations and Data Analysis}

\subsection{Observations}

 We observed PG0946+301 with ASCA (Tanaka \etal 1994) on 1998 November
 12. ASCA contains two sets of two detectors, SIS (Solid-state
Imaging Spectrometer) and GIS (Gas Imaging Spectrometer). The
effective exposure times in SIS0, SIS1, GIS2 and GIS3 were 72,024
seconds, 69,668 seconds, 80,910 seconds and 80,896 seconds
respectively. SIS was operated in 1CCD mode with the target in the
standard 1CCD mode position. GIS was operated in pulse height (PH)
mode. The data were reduced and analyzed using FTOOLS and XSELECT in a
standard manner (see ASCA Data Reduction Guide or MES95 and G99 for
details of data reduction).

\subsection{Image Analysis}

\subsubsection{XSELECT Analysis}

 We used XSELECT to create full and hard (2--9.5 keV) band images of
 for each of the four detectors. We also created combined SIS and GIS
 images. We looked for the target in these images displayed with
 SAOIMAGE. While there were sources seen within the GIS field of view,
 there was no obvious source seen at the target position in any of the
 four detectors. We then smoothed the images with a Gaussian function
 of $\sigma=$ 1--2 pixels. A faint source at the position of the
 target was then evident in GIS3 hard band image and a trace of a
 source was seen in the full GIS image, but not in any other
 image. Note that for a standard pointing position the target lies
 closest to the optical axis in SIS0 and GIS3. GIS3 is more sensitive
 in hard X-rays than SIS0. The fact that the source is seen by eye in
 the GIS3 detector only suggests that the source is faint with flux
 mainly in the hard band.

 We extracted the total counts in a circular region with a
3$^{\prime}$ radius centered on the source position.  Because our
source is observed to be so faint, background subtraction is crucial
in determining the net source count rate, so we have done careful
background subtraction using different background estimates.
Background counts were extracted in two different ways: (1) from a
source-free region on the detector and (2) from exactly the same
region as the source in the blank sky background files provided by the
ASCA guest observer facility.  The significance of the source
detection was therefore different for different background
estimates. For SIS, the blank sky background is underestimated because
it is available in the BRIGHT mode only, while the source counts were
extracted in the BRIGHT2 mode. So the SIS detections are less reliable
with background (2). We found that the source was detected in GIS3 and
GIS3 hard band, and is marginally detected in SIS0 (2$\sigma$).  It
was not detected in any other detector in either bandpass.  The
significance of detection for the source in different detectors and
the resulting net count rate is given in Table 1. For non-detections,
we give a 3$\sigma$ upper limit of the count rate (see G99 for exact
formulation of the detection and corresponding count rate estimate).

\subsubsection{XIMAGE Analysis}

Determination of whether or not the source is detected is extremely
important to our results. As an independent check, we performed image
analysis with XIMAGE (Giommi, Angellini, \& White 1997) which is
designed for detailed image analysis. The $\sf detect$ algorithm in
XIMAGE locates point sources in an image by means of a sliding-cell
method. We used $\sf detect$ on all of our images and looked for a
source at the position of the target. Again, we found the source to be
detected in GIS3 hard band. To minimize the number of spurious sources
detected, the threshold used by $\sf detect$ is somewhat
conservative. As a result, sources with intensity just above the image
background can be missed. We found that the source was detected in the
full band GIS image if we lowered the detection threshold.
 The source was not detected in other detectors. These
results are consistent with those from the XSELECT analysis discussed
above.

\subsubsection{CIAO Analysis}

We applied more sophisticated wavelet-based techniques (Freeman \etal 2000) to
provide
independent support to the above detections.  Software developed for
{\sl Chandra Interactive Analysis of Observations} (CIAO) allows us to
decompose the image such that structures at different scales are
enhanced.  We analyzed the central $20'$ region of GIS3 images in both
the full spectral range and in the harder range.  Wavelet analysis of
the GIS image at scales approximating the size of the point spread
function shows that detection of PG0946+301 is complicated by the
presence of a strong nearby source $\sim 5'$ away.  In the GIS hard
band image, this source is significantly weaker, and we detect
PG0946+301 at a probability of spurious detection of $10^{-4}$, with a
net count rate of $(1.26 \pm 0.25) \times 10^{-3}$ counts s$^{-1}$ (90\%
confidence). This is consistent with the results discussed above.

\subsection{Column Density Constraints}

Consistency among the methods discussed above gives us confidence
in our measurements and in our resulting detections in GIS3 and
non-detections in other detectors. If the low observed X-ray count
rate is due to intrinsic absorption, we can estimate the absorbing
column density in PG0946+301.  Since the source did not yield enough
net counts in any detector to perform spectral analysis, we use the
method discussed in GM96 to determine the column density. We first
calculate the flux from the source if there was no intrinsic
absorption. This was done using the observed $\it B$ magnitude of the source
($\it B=16.0$ mag.) and assuming \aox=1.6.  Redshift of the source (z=1.216)
and
the Galactic column density (N$_H=1.6\times 10^{20}$ atoms cm$^{-2}$,
Murphy \etal 1996) were taken into account to predict the 2--10 keV
flux in the observed band (=7.2$\times 10^{-13}$ erg s$^{-1}$
cm$^{-2}$). A power-law slope with photon index $\Gamma=1.7$ was
used. We then entered this model into the X-ray spectral analysis
software XSPEC (Arnaud 1996), with normalization consistent with the
expected flux and simulated spectra using SIS and GIS response
matrices. The response of the telescope and detectors was taken into
account as well.
The column
density at the redshift of the source was an additional parameter used
in the simulation. If there was no intrinsic absorption, then the
predicted count rate was found to be typically an order of magnitude
larger than the observed one. We then varied the value of the
intrinsic absorption, keeping the normalization constant, until the
predicted and observed column densities matched. The values of
intrinsic column density estimated in this way are given in Table 2.

 This estimate of $N_{\rm H}$ depends upon $\Gamma$ and \aox.  Given the
 observed range of \aox ($\S1$), our adopted value of \aox=1.6 gives
 conservative estimates of column densities. X-ray spectral slopes
 also vary among quasars. So we have estimated N$_H$ for $\Gamma=2.0$
 as well as $\Gamma=1.7$. Flatter spectra  result in even higher
 derived column densities. As shown in Table 2, even the
 conservative estimate results in Thomson thick X-ray absorption in
 PG0946+301, i.e. N$_{\rm H}$\gax$10^{24}$ cm$^{-2}$. The column density
 estimates are consistent with the detection in GIS3 and non detection
 in SIS0.

Alternatively, is it possible that PG0946+301 (and BALQSOs in general)
is intrinsically X-ray weak?  Earlier work (GM96, G99) could not rule
out this possibility. To test this, we estimated the observed SIS0 hard band
count rate for flux consistent with detection in GIS3 hard band, but
no intrinsic absorption. We find that the source would have been
detected in SIS0 hard band at $>8\sigma$ (with $\Gamma=1.7$;
$>7\sigma$ with $\Gamma=2.0$). So we conclude that the observed X-ray
weakness of BALQSOs is due to absorption, and not due to intrinsic
weakness. We cannot, however, rule out the possibility that the source
is intrinsically X-ray weak with an unusual spectral shape
(turning up at around 10 keV, rest frame). It is also possible that
the observed flux is only the scattered component, from a line of
sight different from the absorbing material. This is unlikely in
PG0946+301 which not strongly polarized (Schmidt \& Hines
1999). However, if true, it again implies the existence of X-ray thick
matter along the direct line of sight.

\section{Discussion}

 We have clearly detected the quasar PG0946+301 in our deep ASCA
 observation and we infer that there is Thomson thick X-ray absorption
 ($N_{\rm H}$\gax$10^{24}$ cm$^{-2}$) toward this BALQSO. The use of a
 detection, rather than upper limits, to determine the absorption is
 highly significant. In earlier work, GM96 and G99 had estimated
 absorbing column densities of a few times $10^{22}$ cm$^{-2}$ and
 $10^{23}$ cm$^{-2}$ respectively. However these were based on
 non-detections only and hence yielded only lower limits to the column
 density. A detection provides a much stronger estimate.

 Assuming that there is indeed Thomson thick matter covering the X-ray
 source, can we infer its ionization state? The X-ray absorber will
 cover the optical and UV continuum sources as well, at least
 partially. If the absorber is completely neutral, it will result in
 significant HI opacity, which is not observed (Arav \etal
 1999). If the absorber is completely ionized, then the opacity due to
 Thomson scattering would be the same in the optical, UV and X-rays
 (up to $m_ec^2$).  Thus this scenario by itself cannot account for
 the unusually large values of \aox.  ~If, on the other hand, the
 hydrogen is mostly ionized, but there are still some hydrogen-like
 and helium-like heavy elements, then photoelectric absorption would
 still be the dominant mechanism in X-rays. In the optical/UV, a
 Thomson opacity of one would result in attenuation by a factor of
 2.7. Such attenuation is inferred from polarization studies (Goodrich
 1997, Schmidt \& Hines 1999). The X-ray absorber thus must be at
 least partially ionized and may be responsible for attenuation in the
 optical and UV.

  Whether the X-ray absorber has an ionization state overlapping the
 range of UV BALs and if it outflows with similar velocity remain
 outstanding questions.  It is possible that the X-ray absorber is
 stationary, at the base of winds producing BALs. X-ray continuum
 source might be preferentially covered. X-ray spectroscopy is
 necessary to better probe the nuclear region in BALQSOs. For
 PG0946+301, we predict about 0.015 counts s$^{-1}$ with the XMM PN. A
 reasonable spectrum may be obtained in about 70 ks.





\acknowledgments
We thank K. Arnaud and L. Angelini for help with
XIMAGE.  This work is supported in part by NASA grants NAG5-8360 (PJG,
SM), NAG5-3249 (SM), NAG5-3841 (IS).  The work by W.v.B. at IGPP/LLNL was performed
under the auspices of the US Department of Energy under contract
W-7405-ENG-48.

\clearpage



\clearpage
\thispagestyle{empty}


\begin{table}[h]
\caption{ASCA Count Rates for PG0946+301$^a$ (10$^{-3}$) photons s$^{-1}$}
\begin{tabular}{|lcccccc|}
\tableline\tableline
 & SIS0 & SIS0 hard & SIS1 & GIS2 & GIS3 & GIS3 hard\\
\tableline
Background$^b$ 1 & $0.855 (2\sigma)$ & $<0.74$& $<1.2$ & $<0.75$ &  $1.84
(7.9\sigma)$ & $1.4 (8.7\sigma)$\\
                 & $<1.2$ && $$ & $$ & $$ &\\
Background$^c$ 2 & $1.29 (3\sigma)$ &$0.51 (2\sigma)$& $<1.2$ & $<0.75$ & $0.52
(2\sigma)$ & $0.48 (2\sigma)$\\
                 &  &$<0.74$& $$ & $$ & $<0.83$ & $<0.63$\\
\tableline
\end{tabular}
\small
\noindent
\newline
a. Significance of detection is given in brackets. For non-detections,
3$\sigma$ upper limit is given. For $2\sigma$ detections, $3\sigma$
upper limit is given as well. For SIS0 and GIS3,  hard band count rates are
given as well.
\\
b. With background from a source free
region on the detector. \\
c. With background from blank sky
observations. \\
\end{table}

\begin{table}[h]
\caption{Column Density Constraints ($10^{24}$ atoms cm$^{-2}$) }
\begin{tabular}{|lcccc|}
\tableline\tableline
 Detector & $\Gamma$ & Detection$^a$ & $3\sigma$ Lower Limit$^b$ & $2\sigma$
Detection \\
\tableline
GIS3      &1.7 & 0.95 & 2.1 & 3.3\\
          &2.0 & 0.52 & 1.2 & 1.95 \\
GIS3 hard &1.7 & 1.2 & 2.55 & 3.2\\
          &2.0 & 0.67& 1.55 & 1.95 \\
SIS0      &1.7 &1.4  & 1.42 & 1.95\\
          &2.0 & 0.9 & 0.92 & 1.3 \\
SIS0 hard &1.7 & & 2.12& 2.73 \\
          &2.0 & & 1.42& 1.86 \\
\tableline
\end{tabular}
\small
\noindent
\newline
a. If $3\sigma$ or better detection (Table 1). \\
b. Upper limit on count rate gives lower limit on the column density. \\
\end{table}



\clearpage




\end{document}